# Defining shapes of 2D-crystals with undefinable edge-energies

Luqing Wang, Sharmila N. Shirodkar, Zhuhua Zhang, and Boris I. Yakobson*

Department of Materials Science and NanoEngineering, Rice University, Houston, TX 77005, USA

* biy@rice.edu

**Abstract:** The equilibrium shape of crystals is a fundamental property of both aesthetic appeal and practical import. It is also a visible macro-manifestation of the underlying atomic-scale forces and chemical makeup, most conspicuous in two-dimensional (2D) materials of keen current interest. If the crystal surface/edge energy is known for different directions, its shape can be obtained by geometric Wulff construction, a tenet of crystal physics. However, if symmetry is lacking, the crystal edge energy cannot be defined or calculated, so its shape becomes elusive, presenting an insurmountable problem for theory. Here we show how, in fact, one can proceed with "latent edge energies" towards constructive prediction of a unique crystal shape, and demonstrate it for challenging material-examples, like SnS of $C_{2v}$ symmetry and even $AgNO_2$ of $C_1$–no symmetry at all.

The very word "crystal" we instantly associate with shape (and perhaps color, or the lack of it), often perfected through slow geological formation or craftsmanship. Physical systems in equilibrium arrive to a state of minimal energy. In the case of crystals, oblivious of this fundamental principle, their shapes are achieved by billions of constituent atoms relentlessly performing a trial and error experiment, until they reach the "equilibrium shape". For us to predict a crystal shape, such an approach is impossible, and so theories usually reduce the search to the exterior (surface or edge) energy minimization only[1,2], while the interior-bulk (volume or area) remains invariant. If the exterior energy density $\varepsilon(a)$ is given for all direction-angles $a$, this should suffice for defining the crystal shape, as epitomized by the famed Wulff construction[1,3-5], a geometrical recipe derived from surface energy, where the answer emerges as an envelope of planes or lines distanced by $\varepsilon(a)$ from some point, drawn for all directions $a$.

A century later, the advent of two dimensional (2D) materials[6-9] made such analysis particularly appealing, with daily growing abundance of shape-imagery (easier than characterizing a 3D shape, not to mention improved microscopy): one can learn whether the crystal reached equilibrium or was shaped kinetically, learn about the edge-structures, and about the impact of environment. Furthermore, advances in the first-principles based computations, notably with density functional theory (DFT), nicely complete Wulff construction by offering the $\varepsilon(a)$ with desired accuracy, to predict crystal shape all the way from its elemental-chemical makeup. Such a plan has been successfully realized in numerous cases, when there was a definition for the edge or surface energy. Since the primary well-defined quantity is always the total energy $E_t$, one typically resorts to a ribbon (or slab, in 3D), to define the edge energy (EE) as an excess over the energy of unbounded bulk material $E_b$, roughly as $\varepsilon = (E_t - E_b)/2l$, per length. This works only if the opposite edges are indistinguishable by symmetry, but fails otherwise, yielding a meaningless "average" $\varepsilon$. In some cases, the approach can be augmented by considering a symmetric polygon or polyhedron with all sides identical, as has been realized for 3D GaAs[10], more recently for 2D *h*BN[11] and for metal chalcogenides[12], a broad family[6-8]. This way cannot be taken for granted, since many materials simply lack the symmetry sufficient to design a sample with identical edges (or surfaces). Then the mere definition of surface energy seems to vanish, a disturbing yet simple reality, highlighted by Cahn *et al*. in their remarkable studies[13,14]. The paradox of Wulff construction is that it states how to obtain the shape from a given EE, but the definition of the latter is left out; Cahn *et al.* went further, to show that indeed such definition is fundamentally absent, but did not offer what to do then. Yet we know that nature does find the answer, for each crystal – a true shape. This poses a compelling problem, how to find it in theory?

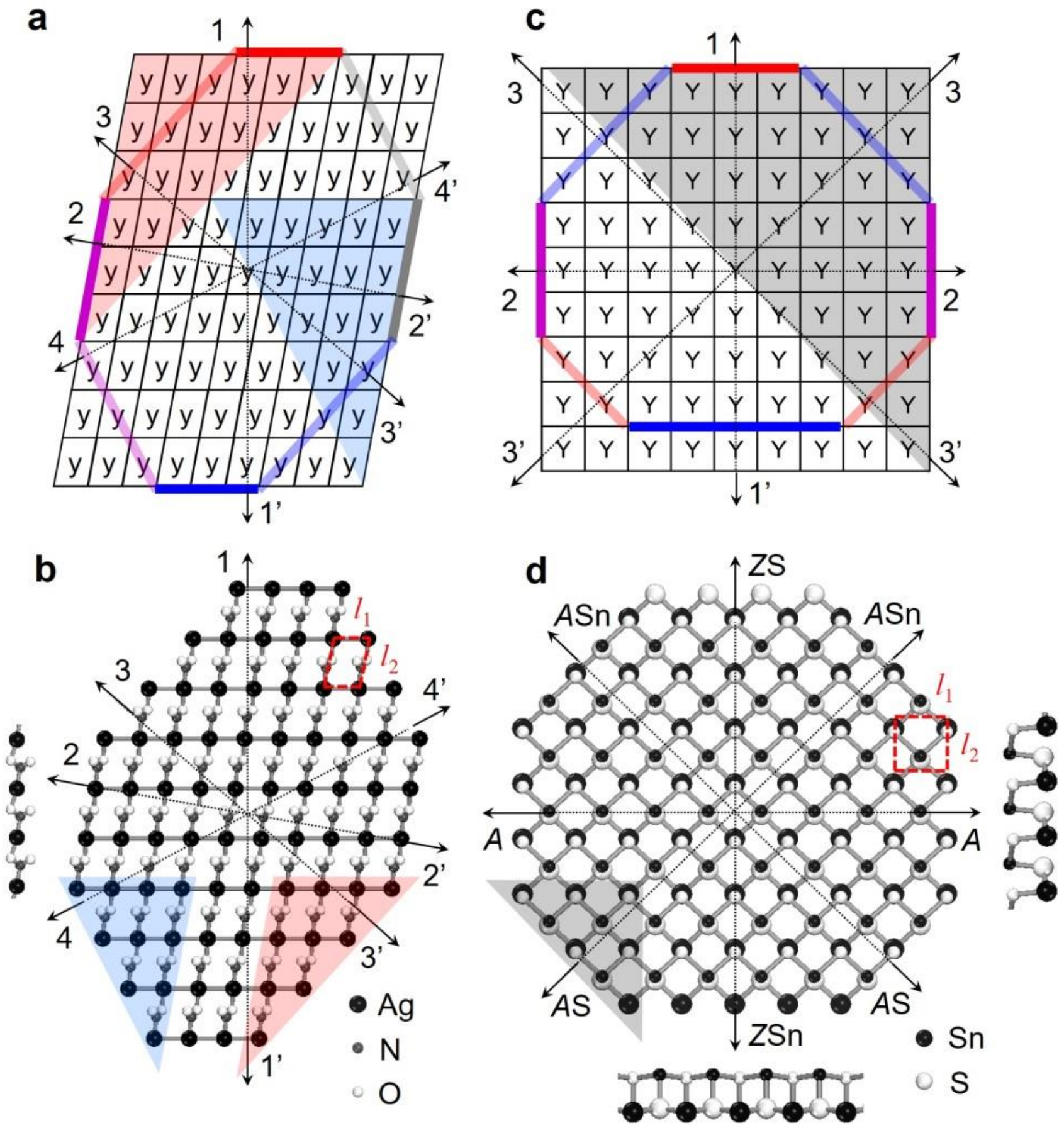

**Fig 1.** The asymmetric 2D-crystals. The "*y*-crystal" (**a**) mimics the $AgNO_2$ monolayer (**b**), with the same lattice constants $l_1$ = 3.39, $l_2$ = 4.93, angle $\angle\alpha = 79.5^o$[15]. Arrows are the normals to 8 basic edges (thick solid lines), red and blue shades mark 2 nonequivalent triangles, and left inset in **b** is the side view. The "**Y**-crystal" (**c**) mimics the SnS monolayer (**d**), with $l_1$ = 4.07, $l_2$ = 4.24, $\angle\alpha = 90^o$[17]. Thick lines highlight 5 basic edges, with their normals as arrows. Small and big atoms in **d** distinguish between top and bottom layers of the SnS, right and bottom insets are side views.

A fully asymmetric ($C_1$) gedanken crystal of "***y***" vividly illustrates such challenge, Fig. 1a: no matter what sample is cut out (ribbon, triangle, circle or any other) it is not surrounded by identical edges, rendering their energies elusive, and the equilibrium shape "unpredictable", at least by the Wulff construct. To be concrete, Fig. 1b shows a fully asymmetric monolayer of silver nitrite[15,16], and a well-studied 2D SnS[17-19] of $C_{2v}$ symmetry (Fig. 1c-d)—slightly higher yet insufficient for separating and defining its EE. Its sketch-depiction (**Y**-crystal, Fig. 1c) has an advantage: not cluttered with atoms and bonds, it clearly displays symmetry and other features of SnS, essential for the compelling problem of finding the shape.

Here we offer a solution, by demonstrating that even for lowest symmetry $C_1$ crystal (i.e., no-symmetry) its shape can be obtained through well-planned calculations, possibly from *ab initio* type, or for that matter, any other atomistic model permitting total energy evaluation. In all cases, the directions can be easily chosen along the Bravais lattice vectors, supplemented by the diagonals (see SI-1), to serve as basic edges, and to construct a general edge energy function $\varepsilon(a)$ for all directions. The total energies of selected polygons allow one to relate the basic EE by linear-algebraic equations which turn out to be underdetermined and entail introducing arbitrary parameters. Nevertheless, as we show, the shapes obtained this way remain invariant (in accord to[13]) to these "latent" parameters' values, and emerge as true equilibrium shapes. We first demonstrate it for $C_{2v}$–symmetry (material as SnS), and then for general no-symmetry $C_1$ case (with $AgNO_2$ as example). We further include the role of chemical potentials in case of binary and ternary compositions, analyze the *h*BN as a test of the proposed method, and describe the symmetry classification (Table S4).

For simplicity, we begin from $C_{2v}$–symmetry materials[20], like SnS (or SnSe[21-23], GeS and GeSe), having only two EE determinable. It has been experimentally observed and extensively studied[17-19]. Its orthorhombic cell and buckled hexagonal lattice with parallel grooves (Fig. 1d), resemble more known phosphorene[9], but are distinguished by the Sn-S bonds off-plane tilt due to different electronegativity of Sn and S. Abstracting from the chemical composition, our Y-crystal is isomorphic to SnS, both having five nonequivalent basic edges marked by their normals in Fig. 1c, with energies $\varepsilon_1$ and $\varepsilon_{1'}$, $\varepsilon_2$, $\varepsilon_3$ and $\varepsilon_{3'}$; the prime notes the inverse directions, so that $\varepsilon_{2'} = \varepsilon_2$ by symmetry.

Wulff construct requires $\varepsilon(a)$ for arbitrary angles $a$, and for this we invoke an ansatz-extrapolation: any slanted, vicinal edge is viewed as a sequence of small segments of the basic edges and, accordingly, its energy is decomposed into a sum of the basic energies, in proper proportions[24-26], like $c_1\varepsilon_1 + c_3\varepsilon_3$, etc. After straightforward trigonometry, one can express the general $\varepsilon(a)$ through basic EEs only and the geometry[24]:

$$\varepsilon(a) = \varepsilon\cdot|\cos(a+C)| \quad (1)$$
$$\varepsilon \equiv (c_1\varepsilon_i^2 + c_2\varepsilon_i\varepsilon_j + c_1\varepsilon_j^2)^{1/2}$$
$$C \equiv \arctan[(c_3 + c_4\varepsilon_j/\varepsilon_i)/(c_5 + c_6\varepsilon_j/\varepsilon_i)]$$

The parameters are specified in the Table S1. We will see soon that the precision of this extrapolation is not essential, having no effect on the results.

Commonly, in simplest situations, the basic EEs are determined (computed) by choosing a sample enclosed by only one edge type: a ribbon for any inversion-symmetry crystal, or an equilateral triangle for a trigonal symmetry case like *h*BN. However, it is impossible to do so for Y-crystal whose symmetry is insufficient. Apart from $\varepsilon_2$ (for which a ribbon can be constructed, see equation 2.2), all other basic edges cannot be singled out by any cutout. Consequently, for *five* unknowns (basic EEs) only *four* independent equations can be set up, using ribbons and triangles (shaded in Fig. 1c) with different edges:

$$\varepsilon_1 + \varepsilon_{1'} = E_{11'}/l_1 \quad (2.1), \qquad \varepsilon_2 = E_{22}/2l_2 \quad (2.2) \qquad \varepsilon_3 + \varepsilon_{3'} = E_{33'}/l_3 \quad (2.3)$$

$$\varepsilon_1 l_1 + \varepsilon_2 l_2 + \varepsilon_{3'} l_3 = E_{123'} \quad (2.4)$$

where $l_i$ are the lattice constants ($l_3 = 5.88$; having the lengths and energies measured in Å and eV, henceforth we omit these units for brevity, without ambiguity). The right hand side (RHS) values are all well-defined, computable total energies of ribbons or triangles (two or three subscripts accordingly), taken relative to the bulk crystal energy, that is $\mu_Y$ (chemical potential of the **Y**-component in its 2D-bulk form). In (2.4) the RHS must be evaluated for larger N-cells wide/tall triangles and then divided by N, normalizing to one cell. For **Y**-crystal as an illustration, we arbitrarily pick reasonable values, e.g. 0.5, 0.1, 0.11 and 0.86 for the RHS of equations (2). Having 5 unknowns but only 4 equations, this linear system is underdetermined, so one cannot obtain the basic EEs and no $\varepsilon(a)$, needed for the Wulff construct. We proceed however by adding a closure-equation, find the crystal shape, and further see that the closure-equation has no influence on the crystal shape, which emerges thus as uniquely defined.

The closure can be in form of constraint on any combination of the basic-EE, e.g. $\varepsilon_3 - \varepsilon_{3'} = \lambda$, for *latent*. Then at each $\lambda$-value the system (2) is solved for the basic $\varepsilon_i$, so that equation (1) gives complete energy function $\varepsilon^{\lambda}(a)$, and the shape of **Y**-crystal is obtained as Wulff plot, Fig. 2a. Remarkably, the tangent lines envelope of the Wulff plot merely translates with $\lambda$, while otherwise giving an invariant well-defined shape. Note that only $\varepsilon_2 = \varepsilon_2$ is physically well defined, $\lambda$-independent, due to mirror symmetry. All others vary broadly, following $\lambda$, but with no effect on crystal shape, and can be dubbed "latent" (like latent heat not changing body temperature). Regarding the ansatz leading to convenient extrapolation (1), it is reassuring to see that the minima of piecewise function $\varepsilon(a)$, essential to the Wulff plot, all correspond to the basic edges; any refinements to the remote "petals" of the $\varepsilon(a)$ in Fig. 2a would not affect the results, i.e. the found shape is robust to approximation used in (1). While for other cases the number of equations (2), their specifics, and the closure may vary, they follow the same structure, which can be called master-system (MS).

Turning now to a factual SnS, one must account for its binary composition. Its five basic edge-directions copy the **Y**-crystal, but now some edges are non-neutral having specific frontier element (like zigzag edge of *h*BN can have either B or N), accordingly labeled: at 0° ($\varepsilon_1$) is zigzag with sulfur *Z*S, at 46° ($\varepsilon_3$) is armchair with tin *A*Sn, at 90° ($\varepsilon_2$) is "neutral" *A*, at 134° ($\varepsilon_{3`}$) – an inversion of $\varepsilon_3$ – is an *A*S, and at 180° ($\varepsilon_{1'}$) – an inversion of $\varepsilon_1$ – is a *Z*Sn.

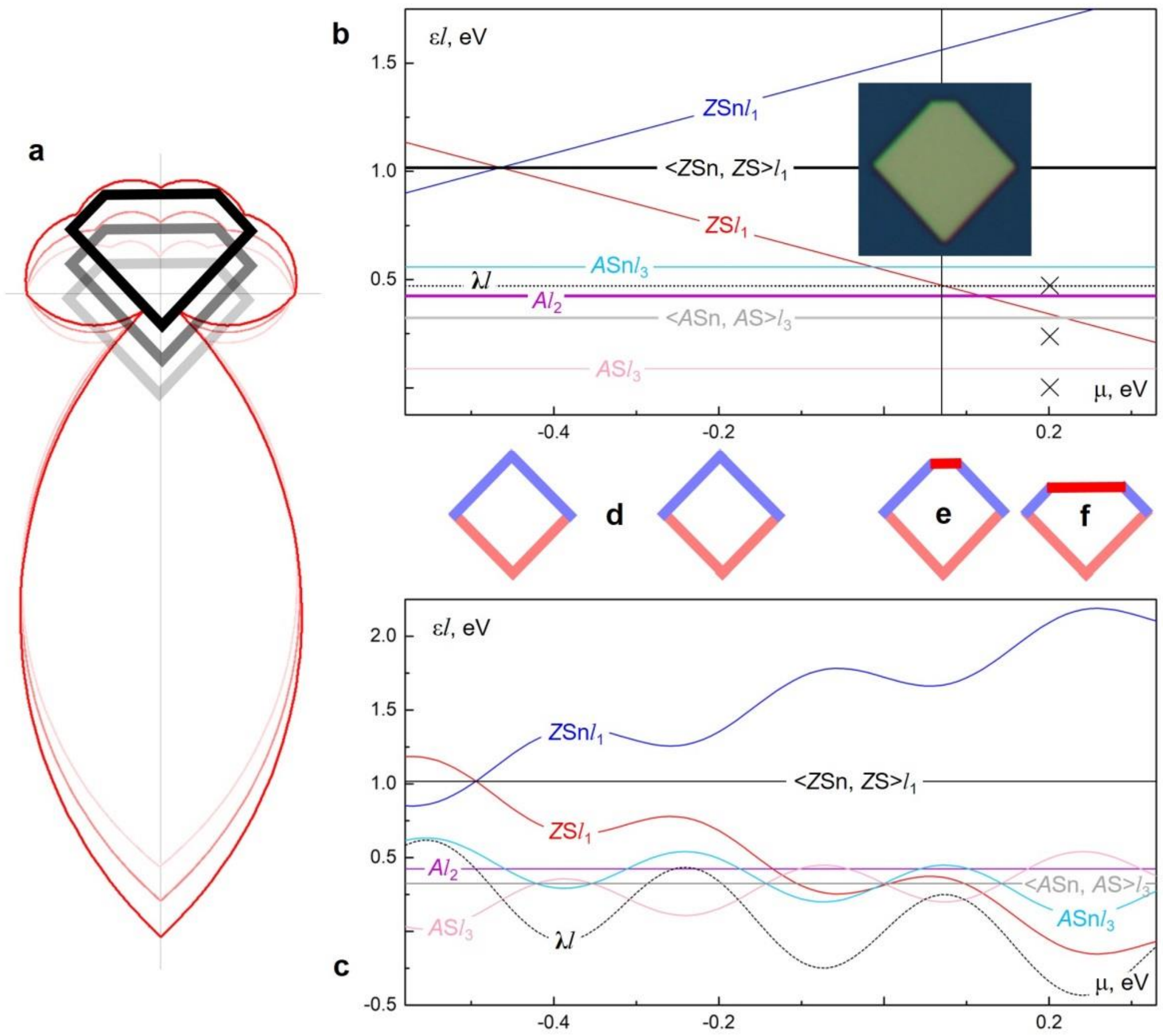

**Fig 2.** The latent edge energies and Wulff constructs of **Y**-crystal and SnS. (**a**) The ε-plots (red) and the Wulff shapes (black) of either **Y**-crystal or SnS, at $\lambda \equiv \varepsilon_3 - \varepsilon_{3'} = 0, 0.04, 0.08$. (**b**) The latent edge energies of SnS varying with chemical potential, at $\lambda \equiv \varepsilon_3 - \varepsilon_{3'} = 0.08$. Black crosses correspond to shapes in **a** or **f**, while the vertical line corresponds to **e** or the experimental image[18] in the inset of **b**. (**c**) The latent EE chosen at random as $\lambda l_3 = -0.59\mu + 0.29 \sin 20\mu$. For the SnS Wulff constructs, $\mu = -0.4$ and $-0.2$ in (**d**), $\mu = 0.07$ in (**e**), and $\mu = 0.2$ in (**f**); Thick red, light-red and light-blue lines represent *Z*S, *A*S and *A*Sn edges, respectively.

The EE function $\varepsilon(a)$ for SnS has the same form (1) as for Y-crystal. Its basic EEs satisfy similar MS like equations (2). The RHS energies can again be taken relative to the bulk crystal energy $\mu_{SnS} = \mu_S + \mu_{Sn}$ (invariant, like $\mu_Y$). The elemental chemical potentials depend on ambient conditions, bringing about new physical variable $\mu \equiv ½ (\mu_S - \mu_{Sn})$. Accordingly, for the triangle 123' having extra sulfur around perimeter, we must include $-\mu$ to the RHS of equation (2.4). For now-specific material, the values in the RHS of equations (2) are obtained from DFT computations of the respective ribbons $E_{11'}/l_1 = 0.50$, $E_{22}/2l_2 = 0.10$, $E_{33'}/l_3 = 0.11$, and triangle $E_{123'} = (1.06 - \mu)$ in equation (2.4). At given conditions, for instance $\mu = 0.2$, we can again complement the algebraic MS with a closure like $\varepsilon_3 - \varepsilon_{3'} \equiv \varepsilon_{ASn} - \varepsilon_{AS} = \lambda$ and compute the crystal shape, Fig. 2a. As we learnt already with **Y**-crystal, the shape stays well defined at a given μ. To reiterate, while the energy of only one edge is certain ($\varepsilon_2 \equiv \varepsilon_A = 0.1$), all others depend on the latent λ, allowed to 'float freely' with no effect on the observable shape. In contrast, μ represents the ambient conditions and can impact the shape. Tracking this is straightforward now: for any μ value assume a λ, and find the EEs versus μ, as plotted in Fig. 2b (with *ad hoc* chosen λ), and then the shapes. Not the individual EEs, but only some combinations are definite: $\varepsilon_{ZS} + \varepsilon_{ZSn}$, $\varepsilon_{AS} + \varepsilon_{ASn}$ are constants (thick lines), while $l_1\varepsilon_{ZS} + l_3\varepsilon_{AS}$ decreases as –μ, at unit slope. Individual EEs however vary with λ, whose choice is arbitrary at each μ, so the functions ε(μ) are merely illustrative (thin lines). To emphasize this, Fig. 2c example shows how unfixed the EEs are, due to the presence of the latent λ chosen as $\lambda(\mu)\cdot l_3 = -0.59\mu + 0.29\cdot\sin 20\mu$; yet the shapes derived from both plots are definite, in Figs. 2d-f. For low μ (Sn-rich side), a rhombus enclosed by *A*Sn and *A*S edges (Figs. 2d) agrees with observed synthetic SnS islands[18]. As μ increases, the shape becomes truncated at one corner (Fig. 2e-f), adding a *Z*S edge. Interestingly, truncated rhombs have also been seen experimentally (Fig. 2b, inset[18]). Together, these facts corroborate the "latent EE" approach to predicting the equilibrium shapes of low symmetry crystals.

Now we turn to the most intriguing, not symmetric at all ($C_1$) ***y***-crystal in Fig. 1a. The 8 basic edges are marked by the normals, with energies $\varepsilon_i$ (i = 1-4, with primes for inverse), and general ε(*a*) has the same form as equation (1) (see Table S2). In absence of any symmetry, the MS extends, relative to equations (2): for the 8 unknowns $\varepsilon_i$ there are 6 relations: 4 with the RHS energies $E_{ii'}$ of the ribbons along all basic directions, plus 2 with the RHS energies $E_{ijk}$ of the triangles, shaded in Fig. 1a (SI, eq. S3). For the abstract ***y***-crystal one simply picks RHS values in the MS, for instance: 0.5, 0.7, 0.6, 0.8 for the ribbons (ii') and 5.1, 5.4 for the triangles (ijk). To be solvable, such underdetermined system must be complemented by *two* closure-conditions, e.g. by assigning arbitrary values (λ, λ') to two of the 8 indeterminate edges, or their combinations. After solving it for all basic EEs, the equation (1) gives $\varepsilon^{\lambda,\lambda'}(a)$ for all directions, to produce the shape of ***y***-crystal using the Wulff plots, Fig. 3a-b. Although the ε-plots vary with latent (λ, λ'), the shape only shifts, remaining the same (inset). This confirms the validity of the "latent EE" method for no-symmetry ($C_1$) case, even with increased number of latencies (here 2, which also is maximum for 2D).

Factual no-symmetry ($C_1$) material example, a monolayer of common salt silver nitrite $AgNO_2$[15,16], has triclinic unit cell, and can be viewed as Ag lattice with $NO_2$ groups inserted between the Ag atoms of the sparse direction of $l_2$, Fig. 1b. The normals of all 8 basic edges are at $a$ = 0°, 48.5°, 79.5°, 117.2°, 180°, 228.5°, 259.5° and 297.2°. For $AgNO_2$, the use of energy expression ε(*a*) for the arbitrarily oriented edge, as well as the MS relating the 8 basic EE, are all like the ***y***-crystal above. New for an *actual* material is that the RHS values in the MS now can be provided as specific DFT-computed values: 0.82, 0.01, 0.52, 0.64 for the ribbons, and near 3.15 for the triangles. Another novelty is in tri-elemental composition, which can still be treated as bi-elemental, Ag and $NO_2$. With $\mu_{Ag} + \mu_{NO2} = \mu_{AgNO2}$ invariant, only one physical parameter must be specified, e.g. chemical potential of silver, $\mu_{Ag}$. It enters the RHS of the MS (SI, eq. S4) in the following way, seen by inspection of Fig. 1b: $\mu_{Ag}$ subtracted from the $E_{11'}$, $E_{33'}$ and $E_{44'}$, for the ribbons naturally containing extra Ag, but not from $E_{22'}$; similarly, for the triangles (shaded in Fig. 1b) one should subtract $\mu_{Ag}$ from both $E_{123'}$ and $E_{12'4}$, to account for the extra Ag, one per primitive cell.

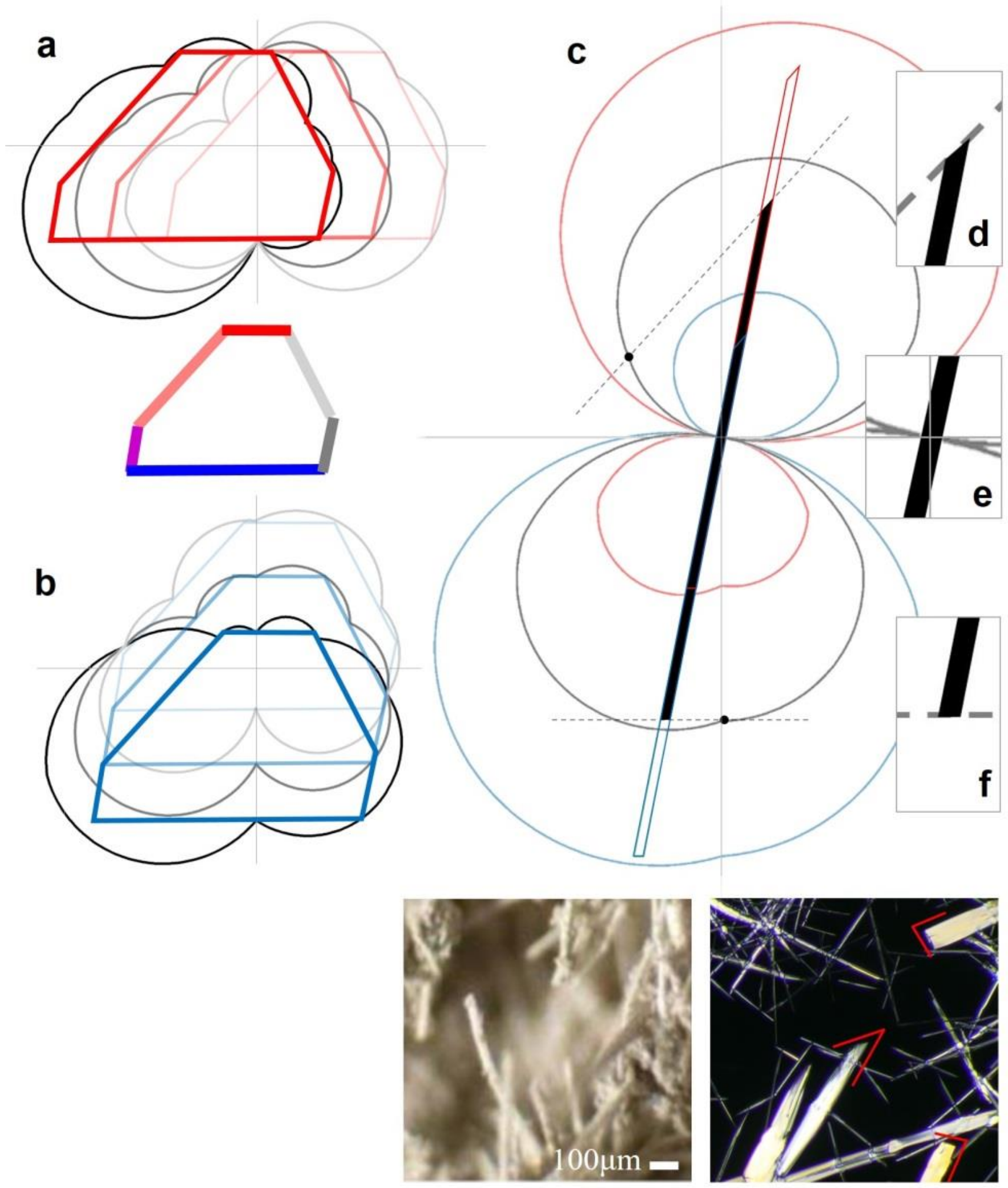

**Fig 3.** The latent ε-plots and Wulff constructions for ***y***-crystal and $AgNO_2$. For ***y***-crystal in (**a**), $(\lambda, \lambda') \equiv (\varepsilon_1 - \varepsilon_{1'}, \varepsilon_2 - \varepsilon_{2'}) = (0, 0.3), (0, 0), (0, -0.3)$; in (**b**), $(\lambda, \lambda') = (-0.3, 0), (0, 0), (0.3, 0)$. Black lines are the ε-plots, red and blue lines are the Wulff shapes, and the inset shows the invariant Wulff shape with edge colors as in Fig. 1a. (**c**) The ε-plot and Wulff constructions for $AgNO_2$ at $\mu_{Ag} = \mu_{Ag\text{-bulk}}$ with blue, grey and red lines for $(\lambda, \lambda') = (-0.42, 0), (-0.02, 0)$ and $(0.38, 0)$. (**d**-**f**) show zoom in of **c**; bottom inset shows experimental images[27], thin red lines highlight the angles at the sample-needle tips, matching well the computed in **d** and **f**.

At a given $\mu_{Ag}$, for some conditions, the MS requires again a closure with two latent parameters, which can be selected as $\lambda = \varepsilon_1 - \varepsilon_{1'}$, $\lambda' = \varepsilon_2 - \varepsilon_{2'}$, in order to proceed and solve now-complete MS of 8 equations, to determine all $\varepsilon_i$ and the entire EE function $\varepsilon(a)$. We choose here not to explore how $\mu_{Ag}$ affects the crystal shape (aspect already covered above for SnS case), but instead assign its value to the bulk silver, and proceed to predict the shape by solving the MS, and then finding the Wulff plots. The results reveal a shape quite extreme and surprising at first, Fig. 3c. To our satisfaction we were able to find a confirmation in rather scarce for $AgNO_2$ experimental evidence (Fig. 3 bottom, insets), where the crystal shapes are quite irregular[27] yet resemble strikingly what theory predicts: highly elongated needle, of no symmetry at all, with one end slanted while another nearly straight.

Having now shown that even for low symmetry, with undefinable EE, the equilibrium crystal shape can be exactly predicted, it is useful to briefly rank all 2D materials in this regard. The very common is (i) trivial-definable case, with inversion symmetry permitting to obtain all EEs directly from total energies of sample-ribbons (e.g. graphene, phosphorene, $SnS_2$). Without it, yet if (ii) a less obvious regular polygon cutout can be found, so we call it nontrivial-definable, all EEs can be unambiguously computed and crystal shape predicted (e.g. *h*BN, $MoS_2$, GaS). Then, among EE-undefinable, there are two levels: (iii) when only a pair of opposite edges permit direct definition while all others remain undefinable (as SnS, SnSe, GeS, GeSe), and (iv) the limit of having no symmetry at all as a foothold (as $AgNO_2$) when indeed none of the EE succumb to definition. In the latter two situations the shape of crystal still can be theoretically predicted, without resort to empirical data, using the "latent EE" approach. We summarize this in a Table S3. We do additional test (SI-6), with a crystal type (ii), like *h*BN, by predicting its shape through the above latent-EE protocol, as if unaware of existing solution based on equilateral triangles; we arrive indeed to identical results.

In summary, predictions or explanations of equilibrium crystal shapes, traditionally done through geometrical Wulff construction, relied on known energy of the surfaces – in case of intensely researched 2D materials, their edges. For materials of low symmetry, the edge energy, however, cannot be computed or even conceptually defined, so one seemingly cannot foresee the shapes without invoking empirical data from experiments[28]. Above we demonstrate how, through a well-planned set of total energy evaluations, augmented by a concept of "latent edge energies", we can restore utility of the Wulff construct and accurately predict the equilibrium shapes of any material. This allowed us to easily include the role of chemical potential, to explore example-materials SnS and fully asymmetric $AgNO_2$, and to predict their shapes (in accord with observations). It is straightforward to generalize this approach to 3D crystals, where our master system (MS) would grow up to 23 linear algebraic equations, plus 3 constraints with latent parameters, still easily solvable for predicting their shapes from first principles. At finite temperatures, one would simply need to replace the RHS of our MS by Gibbs free energies (instead of DFT-computed values) which is well studied and does not interfere with our approach. Presence of substrate can reduce the symmetry of 2D-layers[29], further expanding the necessity of proposed method. Crystals of low symmetry proteins and other biomolecules[30] may also offer broad application to understanding their morphology, beyond the scope of the present work but certainly intriguing.

# Supporting Information

## Defining shapes of 2D-crystals with undefinable edge-energies

Luqing Wang, Sharmila N. Shirodkar, Zhuhua Zhang, and Boris I. Yakobson*

Department of Materials Science and NanoEngineering, Rice University, Houston, TX 77005, USA

* biy@rice.edu

### 1. Discussion on basic edges

We select our basic edges as the primitive lattice vectors $(\bar{l}_1, \bar{l}_2)$, $(-\bar{l}_1, -\bar{l}_2)$and the diagonals $(\bar{l}_1 + \bar{l}_2)$, $-(\bar{l}_1 + \bar{l}_2)$and $(\bar{l}_1 - \bar{l}_2)$, $-(\bar{l}_1 - \bar{l}_2)$. Depending on the symmetry of the material, some of the basic edges become identical and hence their contribution to the edge energy (EE). Though this choice may seem *ad hoc*, we show that chemistry of the material dictates this choice.

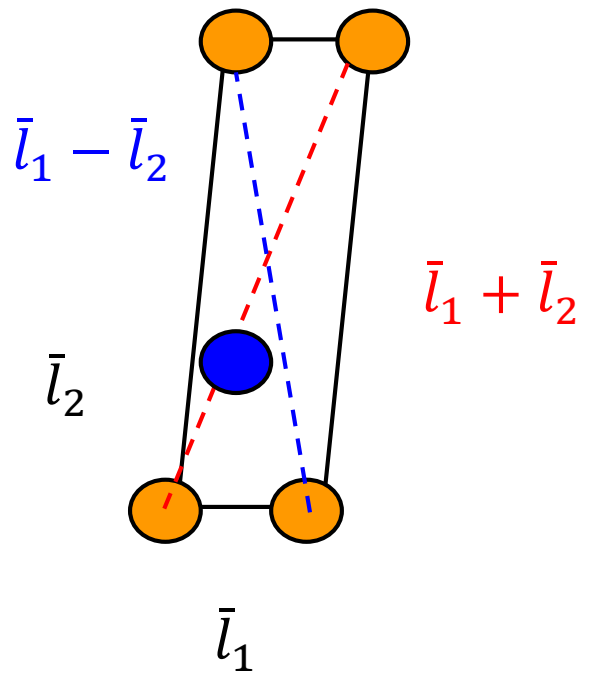

Consider an arbitrary primitive cell with one orange atom in it's basis, as shown here. The distance between atoms along $l_1$ and $l_2$ is smaller than along the diagonals, and hence the contribution to the edge energy due to bonding along these directions is more pronounced. Hence, in this case only the basic edges $(\bar{l}_1, \bar{l}_2)$, $(-\bar{l}_1, -\bar{l}_2)$suffice.

On the other hand, on adding additional basis atom (blue) the bonding along diagonals also becomes important and gives rise to significant contribution to the edge energy. Hence including the diagonals $(\bar{l}_1 + \bar{l}_2)$, $-(\bar{l}_1 + \bar{l}_2)$and $(\bar{l}_1 - \bar{l}_2)$, $-(\bar{l}_1 - \bar{l}_2)$is necessary for the estimation of edge energy.

For eg. in the case of graphene[1] the two basic edges along the primitive lattice vectors are zigzag edges, as seen below in Fig. S1. However, the armchair edge which is along the diagonal has a completely different charge density distribution due to the enhanced bonding between adjacent atoms along this edge. Hence proving that it is important to include the diagonals as the basic edges. Therefore, emphasizing that the EE of an arbitrary directions needs both, zigzag (along primitive cell vectors) and armchair (along diagonal direction) edges, to support the energy decomposition ansatz.

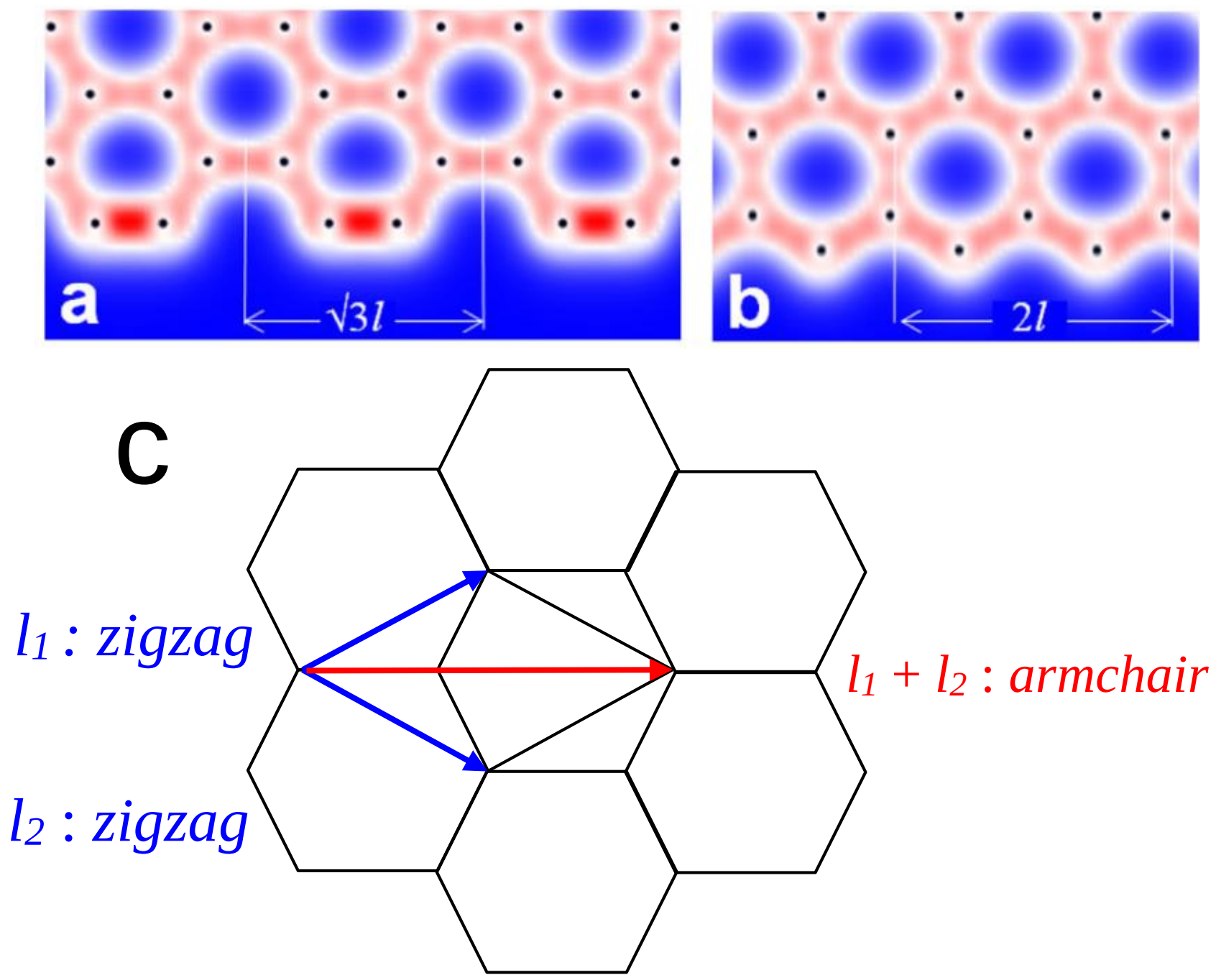

**Figure S1.** Charge density of **(a)** armchair edge and **(b)** zigzag edge in graphene. The armchair edge forms stronger and shorter triple bonds[1]. **(c)** primitive cell of graphene with zigzag edges along primitive lattice vectors and armchair along the diagonal.

## 2. Discussion on tessellation

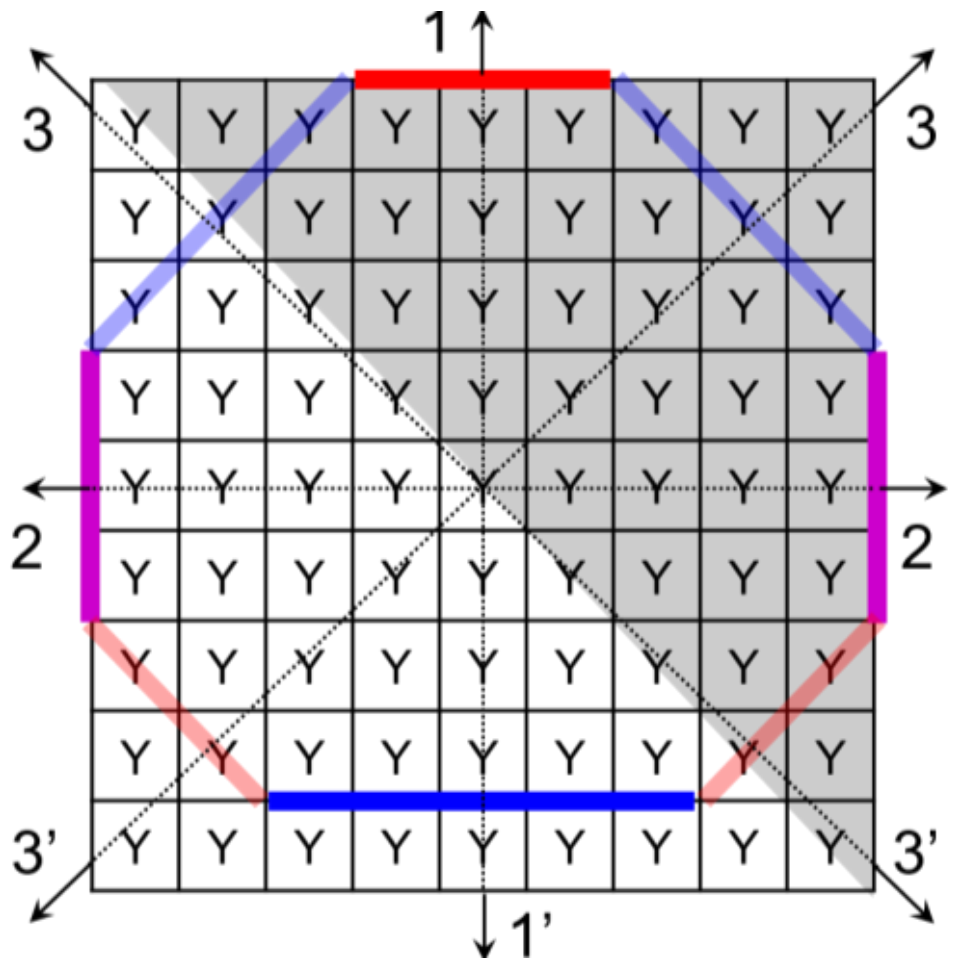

**Figure S2.** Y crystal geometry, with grey shaded triangle.

In Fig S2, the edge energy of grey shaded triangle is
$E_{grey} = \varepsilon_1 + \varepsilon_2 + \varepsilon_{3'}$.

If we add the remaining white triangle to the grey triangle, then the edge energy of combined grey and white polygon is given by:

$E_{grey+white} = \varepsilon_1 + \varepsilon_2 + \varepsilon_{3'} + \varepsilon_2 + \varepsilon_{1'} - \varepsilon_{3'}$

$= \varepsilon_1 + \varepsilon_2 + \varepsilon_{3'} + \varepsilon_2 + \varepsilon_{1'} - \varepsilon_{3'} + \varepsilon_3 - \varepsilon_3$

$= (\varepsilon_1 + \varepsilon_{1'}) + 2\varepsilon_2 + (\varepsilon_3 + \varepsilon_{3'}) - (\varepsilon_3 + \varepsilon_{3'})$.

It reduced to combination of ribbons edge energy. Therefore, beside the basic edge ribbons and the triangle (for "Y" crystal, there're 3 ribbons and 1 triangle), adding other polygons only results in repeating those basic edge ribbons and triangle, and won't introduce new information.

## 3. Cell parameters for the crystals

**3.1** For Y-crystal, the lattice parameters are

$l_1$ = 4.07 Å, $l_2$ = 4.24 Å, and $l_3$ = 5.88 Å (diagonal length)

$\angle\alpha = 90^o$ (angle between $l_1$ and $l_2$)

**3.2** For SnS, the lattice parameters are

$l_1$ = 4.07 Å, $l_2$ = 4.24 Å, and $l_3$ = 5.88 Å (diagonal length)

$\angle\alpha = 90^o$ (angle between $l_1$ and $l_2$)

**3.3** For y-crystal, the lattice parameters are

$l_1$ = 3.39 Å and $l_2$ = 4.93 Å, and $l_3$ = 6.47 Å (along $l_1+l_2$), $l_4$ = 5.46 Å (along $l_1-l_2$)

$\angle\alpha = 79.5^o$ (angle between $l_1$ and $l_2$)

**3.4** For $AgNO_2$, the lattice parameters are

$l_1$ = 3.39 Å and $l_2$ = 4.93 Å, and $l_3$ = 6.47 Å (along $l_1+l_2$), $l_4$ = 5.46 Å (along $l_1-l_2$)

$\angle\alpha = 79.5^o$ (angle between $l_1$ and $l_2$)

## 4. Symbol correspondence between schematic and real crystals

**4.1** Symbols correspondence between Y-crystal and SnS are

1 = *Z*S, 1' = *Z*Sn, 2 = 2' = *A*, 3 = *A*Sn and 3' = *A*S

**4.2** Symbols correspondence between y-crystal and $AgNO_2$ are

1 = 0$^o$, 1' = 180$^o$, 2 = 79.5$^o$, 2' = 259.5$^o$, 3 = 48.5$^o$, 3' = 228.5$^o$, 4 = 117.2$^o$ and 4'= 297.2$^o$

## 5. Master system for calculating total edge energy

**5.1** For Y-crystal the master system is

ribbons

$\varepsilon_1 + \varepsilon_{1'} = 0.5$ eV/Å or $(\varepsilon_1 + \varepsilon_{1'})l_1 = 2$ eV (eq. S1)

$\varepsilon_2 = 0.1$ eV/Å or $\varepsilon_2\, l_2 = 0.42$ eV

$\varepsilon_3 + \varepsilon_{3'} = 0.11$ eV/Å or $(\varepsilon_3 + \varepsilon_{3'})l_3 = 0.65$ eV

triangle

$4.07\varepsilon_1 + 4.24\varepsilon_2 + 5.88\varepsilon_{3'} = 0.86$ eV

Note that, we arbitrarily choose the triangle edge energy for 'Y', remaining energies are computed using DFT for SnS and then used for 'Y'.

**5.2** For SnS the master system is

ribbons

$\varepsilon_{ZS} + \varepsilon_{ZSn} = (E(\text{ribbon})_{11'} - N\cdot\mu_{SnS})/l_1$

$\varepsilon_{AC} = (E(\text{ribbon})_{22} - N\cdot\mu_{SnS})/2l_2$

$\varepsilon_{ASn} + \varepsilon_{ASn} = (E(\text{ribbon})_{33'} - N\cdot\mu_{SnS})/l_3$

S-rich triangle

$M\times(\varepsilon_{ZS}\, l_1 + \varepsilon_A\, l_2 + \varepsilon_{AS}\, l_3) + 3\times E(\triangle\text{vertex}) = E(M\triangle)_{(123')} - N\cdot\mu_{SnS} + N'\cdot\mu_S$

Here, N is the number of stoichiometric bulk units of SnS in the triangle of size M, and N' is the excess of Sn or S for Sn-rich and S-rich triangle respectively. E(ribbon) and E(M△) are the DFT energies of ribbons and triangle of size M, respectively, and E(△vertex) is the energy contribution from the vertex of the triangle. Note that, the vertex energy contribution is constant with size M, and at small sizes, edge and vertex contributions dominate the total energy of triangle E(M△), giving an inaccurate estimate. Hence, one needs to check the scaling of edge energies with increasing M, to get a reasonable estimate for the same. The results of this convergence are plotted in Fig. S2 below. Detailed discussion on this can be found in Y. Liu *et al*.[2]

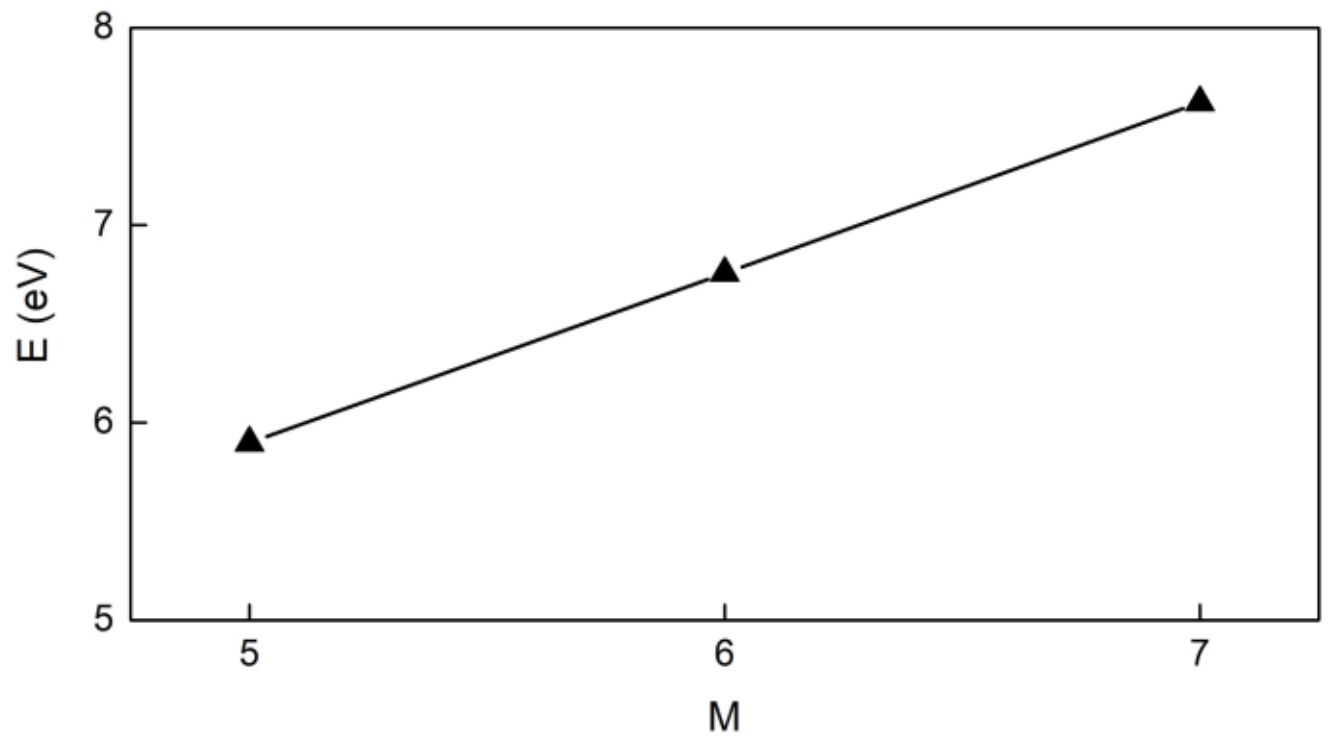

**Figure S3.** Variation of $[E(M\triangle)_{(1'23)} - N\cdot\mu_{SnS} - N'\cdot\mu_S]$ in S-rich triangle in SnS for $\mu = 0.2$ eV. The energy scales linearly with size M.

Though there exist 5 equations and 5 unknowns in the above DFT energy equations, no all are linearly independent. We find that, sum of ribbon equations = sum of triangle equations. Hence, we have 5 unknowns and 4 linearly independent equations. Below is the final MS after substitution of DFT energies for ribbons and triangles.

ribbons

$\varepsilon_{ZS} + \varepsilon_{ZSn} = 0.5$ eV/Å or $(\varepsilon_{ZS} + \varepsilon_{ZSn})l_1 = 2$ eV (eq. S2)

$\varepsilon_A = 0.1$ eV/Å or $\varepsilon_A l_2 = 0.42$eV

$\varepsilon_{ASn} + \varepsilon_{AS} = 0.11$ eV/Å or $(\varepsilon_{ASn} + \varepsilon_{AS})l_3 = 0.65$ eV

triangle

$4.07\varepsilon_{ZS} + 4.24\varepsilon_A + 5.88\varepsilon_{AS} = (1.06 - \mu)$ eV (S-rich triangle)

Note that, we only change the triangle edge energy for 'Y', remaining energies are computed using DFT for SnS and then used for 'Y'.

**5.3** For y-crystal the master system is

ribbons

$\varepsilon_1 + \varepsilon_{1'} = 0.5$ eV/Å (eq. S3)

$\varepsilon_2 + \varepsilon_{2'} = 0.7$ eV/Å

$\varepsilon_3 + \varepsilon_{3'} = 0.6$ eV/Å

$\varepsilon_4 + \varepsilon_{4'} = 0.8$ eV/Å

triangles

$3.39\varepsilon_1 + 4.93\,\varepsilon_2 + 6.47\,\varepsilon_{3'} = 5.1$ eV

$3.39\,\varepsilon_1 + 5.46\,\varepsilon_4 + 4.93\,\varepsilon_{2'} = 5.4$ eV

Note that, the (right hand side) RHS values for the above MS equations have been chosen arbitrarily.

**5.4** For $AgNO_2$ the master system is

Definition of $\mu$

$\mu_{Ag} + \mu_{NO2} = \mu_{AgNO2}$

Fix $\mu_{Ag} = \mu(\text{Ag bulk}) = -2.81$ eV

ribbons

$l_1\,(\varepsilon_1 + \varepsilon_{1'}) = E(\text{ribbon})_1 - 2\,\mu_{AgNO2} - \mu_{Ag}$

$l_2\,(\varepsilon_2 + \varepsilon_{2'}) = E(\text{ribbon})_2 - 4\,\mu_{AgNO2}$

$l_3\,(\varepsilon_3 + \varepsilon_{3'}) = E(\text{ribbon})_3 - 5\,\mu_{AgNO2} - \mu_{Ag}$

$l_4\,(\varepsilon_4 + \varepsilon_{4'}) = E(\text{ribbon})_4 - 4\,\mu_{AgNO2} - \mu_{Ag}$

triangles

$(6 - 5) \times (l_1\,\varepsilon_1 + l_2\,\varepsilon_{2'} + l_4\,\varepsilon_4) = E(6\triangle)_a - E(5\triangle)_a - 7\,\mu_{AgNO2} - \mu_{Ag}$

$(6 - 5) \times (l_1\,\varepsilon_1 + l_2\,\varepsilon_2 + l_3\,\varepsilon_{3'}) = E(6\triangle)_b - E(5\triangle)_b - 7\,\mu_{AgNO2} - \mu_{Ag}$

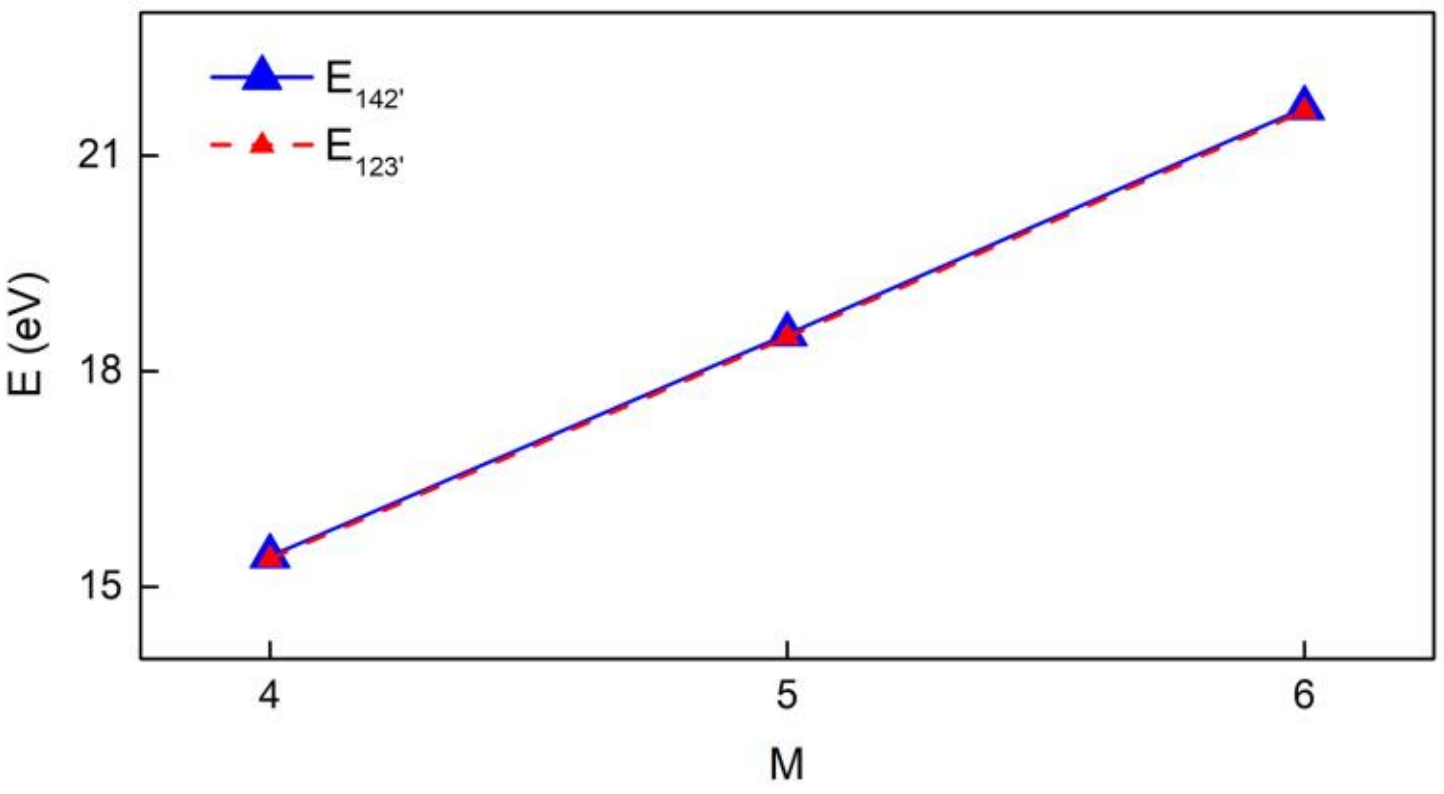

**Figure S4.** Variation of [E(M△) – N $\mu_{AgNO2}$ – N' $\mu_{Ag}$] in $AgNO_2$, for both a (142') and b(123') triangles for $\mu_{Ag}$= –2.81 eV. The energy scales linearly with size M.

Below is the final master system after substitution of DFT energies for ribbons and triangles at $\mu_{Ag}$ = –2.81 eV.

ribbons

$\varepsilon_1 + \varepsilon_{1'}$ = 0.82 eV/Å (eq. S4)

$\varepsilon_2 + \varepsilon_{2'}$ = 0.01 eV/Å

$\varepsilon_3 + \varepsilon_{3'}$ = 0.52 eV/Å

$\varepsilon_4 + \varepsilon_{4'}$ = 0.64 eV/Å

triangles

$3.39\varepsilon_1 + 4.93\ \varepsilon_2 + 6.47\ \varepsilon_{3'}$ = 3.15 eV

$3.39\ \varepsilon_1 + 5.46\ \varepsilon_4 + 4.93\ \varepsilon_{2'}$ = 3.15 eV

## 6. Wulff shapes of BN

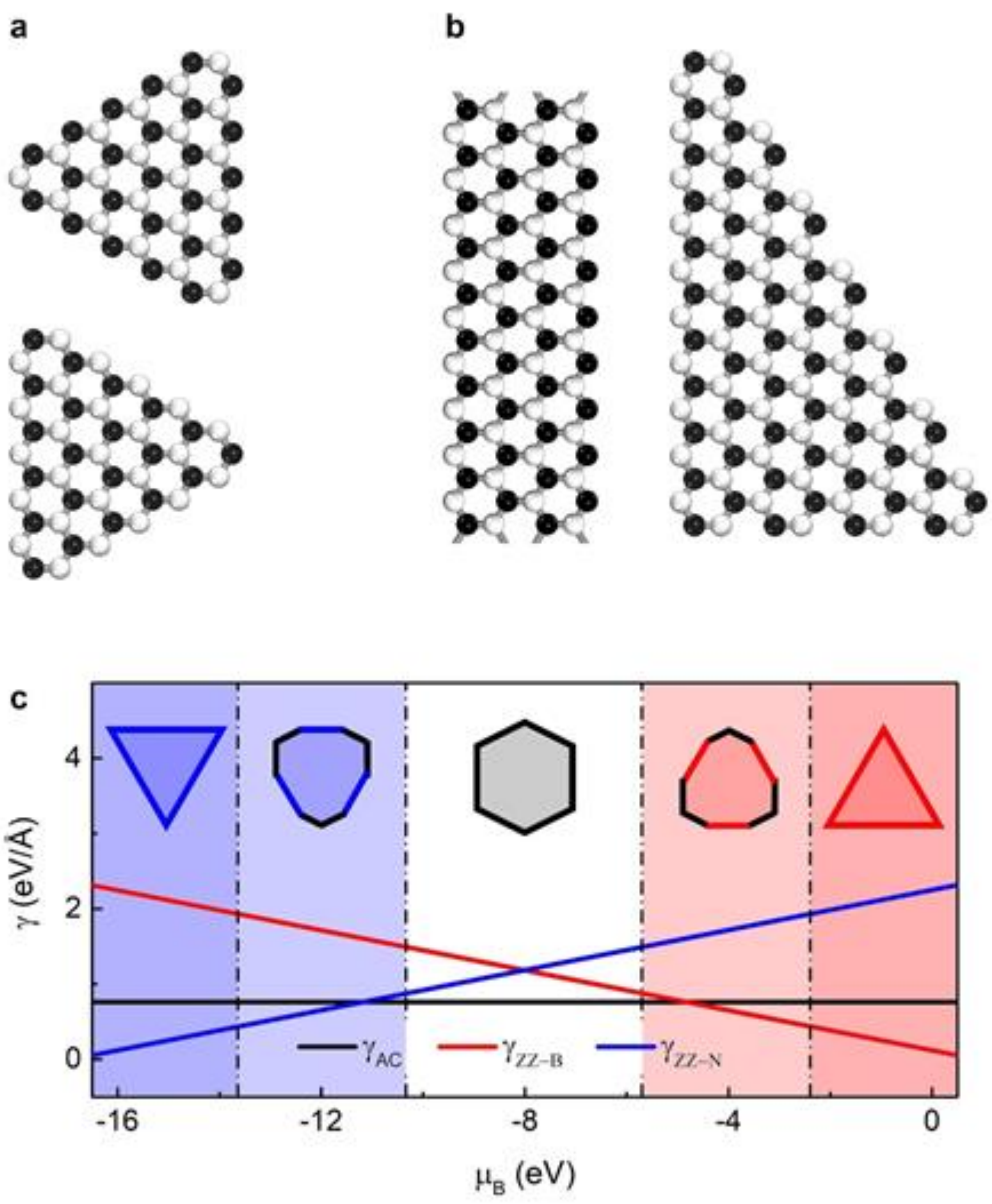

**Figure S5.** Structure, edge energies and ES of BN. **(a)** BN triangles with identical edges used in previous work. **(b)** BN ribbon and triangle with different edges used in this work. **(c)** edge energies of AC, ZZ-B and ZZ-N edges as chemical potential of sulfur $\mu_B$ varies. Inserted polygons are ES corresponding to different $\mu_B$ range, which are separated by dot dash line and shaded in different color, respectively.

As established from our analysis and examples, the "latent edge energy" method can determine the ES of 2D low-symmetry materials with no-well defined edge energy; where some edges can never be isolated and hence their energies never resolved. We note that a situation when particle is in contact with a substrate leads to additionally reduced symmetry, but the treatment has always been fully based on fully known surface and interface (contact) energies. This approach, often known as Winterbottom[3] or Wulff-Kaischew[4] construction, is just a minor modification of the Wulff theorem[5]. It also extends the traditional Wulff construction use by overcoming the limitation that the ES can only be determined for materials with separable edges. To establish the validity and gain new perspective into the relationship between "latent edge energy" method and traditional ES determining techniques, we apply it to sufficient symmetry materials for which the solution for ES is readily available[7]. Here, we apply our method to monolayer hexagonal boron nitride (BN).

It's well-known that BN has $D_{3h}$ symmetry, and three basic directions, A, Z-B and Z-N. The edge energy $\varepsilon_{BN}(a)$ along an arbitrary orientation can be expressed as eq. S1:

$\varepsilon(a) = |\varepsilon| \cos(a + C)$

where $|\varepsilon| = 2\,(\varepsilon_A^2 + \varepsilon_{Zx}^2 - \sqrt{3}\,\varepsilon_A\,\varepsilon_{Zx})^{1/2}$ and $C = \mathrm{sgn}(a) \cdot \arctan(\sqrt{3} - 2\,\varepsilon_{Zx}/\varepsilon_A)$ (eq. S5)

with the subscript $x$ = N or $x$ = B. Since Z-B and Z-N edges can't be separated by a ribbon, they were in previous work calculated using an equilateral triangle with identical edges[7] (Figure S3a). However, in the "latent edge energy" method the determination of $\varepsilon_{basic}$ skips the construction of polygons with identical edges, i.e. it overcomes the necessity of isolating all the edges. Instead, we use a ribbon with opposite Z-N and Z-B edges, as seen in Figure S3b. On the other hand, the A edge can be isolated with a ribbon, and $\varepsilon_A$ = 0.76 eV/Å is estimated (in agreement with previous work[7].

Though the A edge is independent of μ, the dependence of Z edges on μ affects the delicate balance between $\varepsilon_{basic}$, which changes the ES of BN. In order to establish the relation between $\varepsilon_{Z\text{-}B}$, $\varepsilon_{Z\text{-}N}$ and $\mu_B$ (chemical potential of B), we construct a triangle with *non-identical edges*, an advantage of the "latent energy method", as shown in Supplementary Information Figure S3b. We arrive at the same expressions as Liu *et al.*[7], $\varepsilon_{Z\text{-}B} = (0.12 - 0.13\mu_B)$ eV/Å and $\varepsilon_{Z\text{-}N} = (0.12 + 0.13\mu_B)$ eV/Å. These $\varepsilon_{basic}(\mu_B)$ when substituted in equation for ε, reproduce ES of BN as a function of $\mu_B$ as shown in Figure S3c (in agreement with previous results[7]. We find that from low to high $\mu_B$, ES evolves from triangle with Z-N edges → enneagon with Z-N and A edges → hexagon with A edges → enneagon with Z-B and A edges → triangle with Z-B edges. Thus, confirming the validity and applicability of our "latent edge energy" method.

## 7. DFT Methods

To obtain numerical values for specific materials, as the RHS of the MS, like in equation (2), the first-principles density functional theory (DFT) calculations and structural optimization were performed using the Vienna Ab initio Simulation Package (VASP)[6,7], adopting generalized gradient approximation (GGA) with the Perdew-Burke-Ernzerhof (PBE)[8] exchange-correlation functional along with the projector-augmented wave (PAW) potentials. Electronic wave functions were expanded in a plane wave basis set with the kinetic energy cutoff of 400 eV and for the Brillouin zone integration a 9 × 1 × 1 Monkhorst-Pack k-point mesh was used for ribbon structures. The energy convergence criterion for electronic wave function was set to be $10^{-5}$ eV. A vacuum layer of about 10 Å in z direction was chosen to guarantee negligible spurious interaction between layers in monolayer simulations using periodic boundary conditions.

**Table S1.** The parameters of equation (1) for Y-crystal (and SnS).

| $a$ | $\varepsilon_i$ | $\varepsilon_j$ | $c_1$ | $c_2$ | $c_3$ | $c_4$ | $c_5$ | $c_6$ |
|---|---|---|---|---|---|---|---|---|
| $0^o \leq a < 46^o$ | $\varepsilon_1$ ($\varepsilon_{ZS}$) | $\varepsilon_3$ ($\varepsilon_{ASn}$) | 1.92 | -2.66 | 0.96 | -1.39 | 1 | 0 |
| $46^o \leq a < 90^o$ | $\varepsilon_3$ ($\varepsilon_{ASn}$) | $\varepsilon_2$ ($\varepsilon_A$) | 2.09 | -3.01 | 0 | 1 | -1.44 | 1.04 |
| $90^o \leq a < 134^o$ | $\varepsilon_2$ ($\varepsilon_A$) | $\varepsilon_{3'}$ ($\varepsilon_{AS}$) | 2.09 | -3.01 | 1 | 0 | -1.04 | 1.44 |
| $134^o \leq a < 180^o$ | $\varepsilon_{3'}$ ($\varepsilon_{AS}$) | $\varepsilon_1$ ($\varepsilon_{ZSn}$) | 1.92 | -2.66 | 1.39 | -0.96 | 0 | 1 |

**Table S2.** The parameters of equation (1) for y-crystal (and $AgNO_2$).

| $a$ | $\varepsilon_i$ | $\varepsilon_j$ | $c_1$ | $c_2$ | $c_3$ | $c_4$ | $c_5$ | $c_6$ |
|---|---|---|---|---|---|---|---|---|
| $0^o \leq a < 48.5^o$ | $\varepsilon_3$ | $\varepsilon_1$ | 1.78 | -2.36 | -1.335 | 0.885 | 0 | 1 |
| $48.5^o \leq a < 79.5^o$ | $\varepsilon_2$ | $\varepsilon_3$ | 3.77 | -6.46 | -1.287 | 0.354 | -1.454 | 1.909 |
| $79.5^o \leq a < 117.2^o$ | $\varepsilon_4$ | $\varepsilon_2$ | 2.67 | -4.23 | 0.298 | 0.748 | 1.609 | -1.454 |
| $117.2^o \leq a < 180^o$ | $\varepsilon_4$ | $\varepsilon_{1'}$ | 1.26 | -1.16 | 1.125 | -0.514 | 0 | 1 |
| $180^o \leq a < 228.5^o$ | $\varepsilon_{3'}$ | $\varepsilon_{1'}$ | 1.78 | -2.36 | -1.335 | 0.885 | 0 | 1 |
| $228.5^o \leq a < 259.5^o$ | $\varepsilon_{2'}$ | $\varepsilon_{3'}$ | 3.77 | -6.46 | -1.287 | 0.354 | -1.454 | 1.909 |
| $259.5^o \leq a < 297^o$ | $\varepsilon_{4'}$ | $\varepsilon_{2'}$ | 2.67 | -4.23 | 0.298 | 0.748 | 1.609 | -1.454 |
| $297^o \leq a < 360^o$ | $\varepsilon_{4'}$ | $\varepsilon_1$ | 1.26 | -1.16 | 1.125 | -0.514 | 0 | 1 |

**Table S3.** The types of 2D materials according to the determinacy of $\varepsilon$.

| The determinacy of $\varepsilon$ | Inversion symmetry | The number of directions of solvable $\varepsilon$ |
|---|---|---|
| well defined | Yes | >2 |
| | No | >2 |
| no-well defined | No | 2 |
| | No | 0 |

**Table S4.** The classification of 230 space groups according to no-well defined edge energy problem. Here x, y, z are directions normal to the 2D plane, and the 3D space groups belonging to the point groups are given in the brackets.

| | x | y | z |
|---|---|---|---|
| $W_i$ | $C_i$(2), $C_{2h}$(10-15), $D_{2h}$(47-74), $C_{4h}$(83-88), $D_{4h}$(123-142), $S_6$(147, 148), $D_{3d}$(162-167), $D_{6h}$(191-194), $T_h$(200-206), $O_h$(221-230) | $C_i$(2), $C_{2h}$(10-15), $D_{2h}$(47-74), $C_{4h}$(83-88), $D_{4h}$(123-142), $S_6$(147, 148), $D_{3d}$(162-167), $D_{6h}$(191-194), $T_h$(200-206), $O_h$(221-230) | $C_i$(2), $C_{2h}$(10-15), $D_{2h}$(47-74), $C_{4h}$(83-88), $D_{4h}$(123-142), $S_6$(147, 148), $D_{3d}$(162-167), $D_{6h}$(191-194), $T_h$(200-206), $O_h$(221-230) |
| $W_n$ | $D_2$(16-24), $D_4$(89-98), $D_{2d}$(111-116, 119-122), $D_3$(150, 152, 154, 155), $D_{3h}$(187-190), T(195-199), O(207-214), $T_d$(215-220) | $C_2$(3, 5), $D_2$(16-24), $D_4$(89-98), $D_{2d}$(111-116, 119-122), $D_3$(150, 152, 154, 155), $D_{3h}$(187-190), T(195-199), O(207-214), $T_d$(215-220) | $D_2$(16-24), $C_{2v}$(25-46), $C_4$(75, 77, 79), $S_4$(81, 82), $D_4$(89-98), $C_{4v}$(99-110), $D_{2d}$(111-122), $C_3$(143, 146), $D_3$(149-155), $C_{3v}$(156-161), $C_6$(168, 171-173), $C_{3h}$(174), $C_{6v}$(183-186), $D_{3h}$(187-190), T(195-199), O(207-214), $T_d$(215-220) |
| $N_2$ | $C_2$(3-5), $C_s$(6-9), $C_{2v}$(25-46), $C_4$(75-80), $S_4$(81, 82), $C_{4v}$(99-110), $D_{2d}$(117, 118), $D_3$(149, 151, 1535), $C_{3v}$(156, 158, 160, 161), $C_6$(168-173), $C_{3h}$(174), $C_{6v}$(183-186) | $C_{2v}$(25-46), $C_4$(75-80), $S_4$(81, 82), $C_{4v}$(99-110), $D_{2d}$(117, 118), $D_3$(149, 151, 153), $C_{3v}$(156, 158, 160, 161), $C_6$(168-173), $C_{3h}$(174), $C_{6v}$(183-186) | $C_2$(3-5), $C_s$(6-9) |
| $N_0$ | $C_1$(1), $C_3$(143-146), $C_{3v}$(157, 159) | $C_1$(1), $C_2$(4), $C_s$(6-9), $C_3$(143-146), $C_{3v}$(157, 159) | $C_1$(1), $C_4$(76, 78, 80), $C_3$(144, 145), $C_6$(169, 170) |